\documentclass[journal]{IEEEtran}

\usepackage[english]{babel}
\usepackage{amssymb}
\usepackage{framed}
\usepackage{array}
\usepackage{enumitem}
\usepackage{amsmath}
\usepackage{amsfonts}
\usepackage{amssymb}
\usepackage{wrapfig,lipsum,booktabs}
\usepackage{graphicx}
\usepackage{times}
\usepackage{booktabs}
\usepackage{t1enc}
\usepackage{amsmath,amsfonts,graphicx}
\usepackage{float}
\usepackage{epstopdf}
\usepackage{hyperref}
\usepackage{comment}
\usepackage{geometry}
\usepackage[export]{adjustbox}

\begin{document}
\bstctlcite{IEEEexample:BSTcontrol}
\title{Identification of Errors-in-Variables ARX Models Using Modified Dynamic Iterative PCA}

\author{Deepak~Maurya, ~Arun~K.~Tangirala,~and~Shankar~Narasimhan
\thanks{Deepak Maurya is currently a research scholar at the Department of Computer Science, Indian Institute of Technology Madras (IIT Madras), India, email: ee11b109@ee.iitm.ac.in, web: \href{https://d-maurya.github.io/}{https://d-maurya.github.io/}}
\thanks{Arun K. Tangirala is a professor at the Department of Chemical Engineering, IIT Madras, India, email: arunkt@iitm.ac.in, web:  \href{http://arunkt.wixsite.com/homepage}{http://arunkt.wixsite.com/homepage}}
\thanks{Shankar Narasimhan is a professor at the Department of Chemical Engineering, IIT Madras, India, email: naras@iitm.ac.in, web: \href{http://www.che.iitm.ac.in/~naras/}{http://www.che.iitm.ac.in/~naras/}}
}

\maketitle

\begin{abstract}
Identification of autoregressive models with exogenous input (ARX) is a classical problem in system identification. This article considers the errors-in-variables (EIV) ARX model identification problem, where input measurements are also corrupted with noise. The recently proposed DIPCA technique solves the EIV identification problem but is only applicable to white measurement errors. We propose a novel identification algorithm based on a modified Dynamic Iterative Principal Components Analysis (DIPCA) approach for identifying the EIV-ARX model for single-input, single-output (SISO) systems where the output measurements are corrupted with coloured noise consistent with the ARX model. Most of the existing methods assume important parameters like input-output orders, delay, or noise-variances to be known. This work's novelty lies in the joint estimation of error variances, process order, delay, and model parameters. The central idea used to obtain all these parameters in a theoretically rigorous manner is based on transforming the lagged measurements using the appropriate error covariance matrix, which is obtained using estimated error variances and model parameters.  Simulation studies on two systems are presented to demonstrate the efficacy of the proposed algorithm.
\end{abstract}

\begin{IEEEkeywords}
error in variables, ARX model, spectral decomposition, model identification,  principal component analysis
\end{IEEEkeywords}

\IEEEpeerreviewmaketitle

\section{Introduction}
\begin{figure}[thpb]
      \centering
      \includegraphics[scale=0.41]{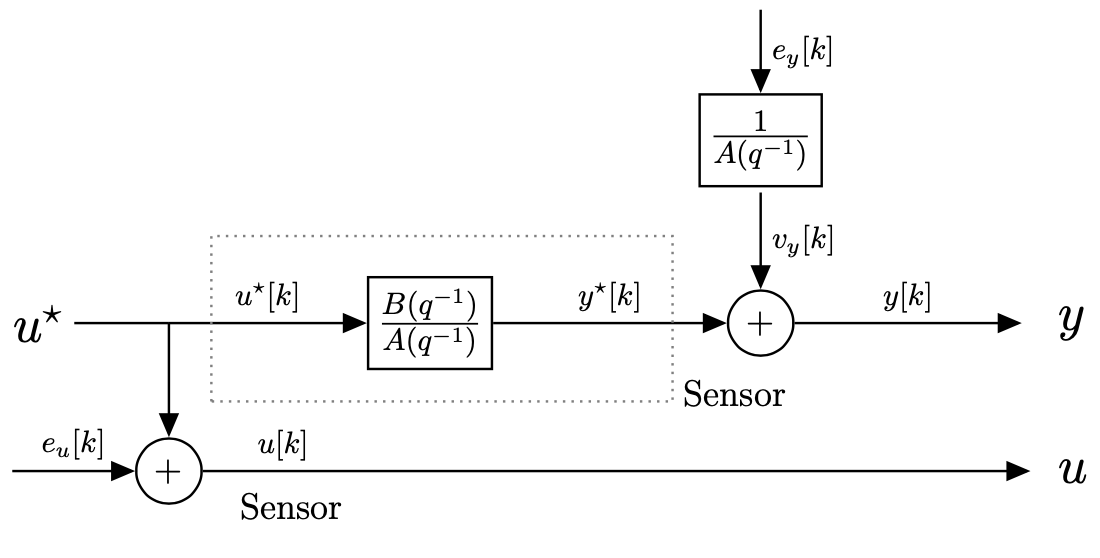}
      \caption{Linear Dynamic EIV ARX Model Structure}
      \label{fig:eiv_arx}
\end{figure}

\IEEEPARstart{S}ystem identification is widely used across various engineering disciplines for developing models from data.  These models are subsequently used for different applications such as control \cite{jin2005frequency} or fault diagnosis \cite{forrai2016system, palmer2018active}.
Identification of autoregressive models with exogenous input (ARX) models is one of the classical problems in system identification \cite{book:ljung}. Several algorithms have been proposed for identifying ARX models using noise-free inputs.  However, in many practical cases, the inputs are also corrupted with noise. Identifying models using noisy inputs is also known as errors-in-variables (EIV) identification \cite{book:soderstorm_eiv}. The discrete system architecture of a single-input single-output (SISO) EIV-ARX process is shown in Figure \ref{fig:eiv_arx}.

In Figure \ref{fig:eiv_arx}, $q^{-1}$ is the usual difference operator, and $u^{\star}[k]$, $y^{\star}[k]$ are the noise-free input and output, respectively, for the $k^{\text{th}}$ sample.  The system transfer function is given by  $\frac{B(q^{-1})}{A(q^{-1})}$. The input measurements are assumed to be corrupted by white noise, while the output measurements are corrupted by colored noise. In the ARX structure, the noise filter corrupting the output, and the system transfer function share the same denominator. The problem is to identify all transfer function parameters and noise variances using $N$ samples of the noisy measurements, $y[k]$, and $u[k]$. 

It is well known that the ordinary least squares method produces biased estimates for the EIV problem \cite{van2004total,soderstrom1981identification}. Several methods have been developed to obtain consistent estimates for the EIV model parameters.  They include methods such as maximum likelihood estimation \cite{diversi2007maximum}, instrumental variable \cite{soderstrom2011generalized,stoica1995combined}, bias compensation \cite{ikenoue2005, zheng2002bias},  Koopmans–Levin  \cite{fernando1985identification}, and recursive estimation \cite{chen2007recursive}. These methods have also been  extended for multi-input and multi-output systems \cite{diversi2017frisch,liu2018arx}. A comprehensive survey on the estimation of the EIV models is described in \cite{book:soderstorm_eiv, soderstrom2007errors}. All of these methods make one or more of the following restrictive assumptions:
\begin{enumerate}
    \item Output noise is white and not colored. 
    \item Noise variances of input and output measurements are known or their ratio is  known. 
    \item Input-output delay \& orders are known.
\end{enumerate}
Among the approaches proposed for EIV model identification, the method based on dynamic Frisch scheme \cite{diversi2010identification}, and the bias compensation or bias elimination methods \cite{ikenoue2005} estimate input and output error variances (which may not be identical) along with ARX model parameters. However, all these methods assume that the process order is known. Thus, there is no single method currently available for EIV-ARX model identification which  is capable of estimating the process order, error variances and model parameters simultaneously.  Even the methods which assume that the order of the process is unknown, attempt to estimate it in a heuristic manner using one of the following approaches:
\begin{enumerate}
    \item \textit{Pre-Estimation}: In this class of approaches, a user estimates the model order and delay before the model estimation. Some of the widely accepted approaches employ step-response analysis,  frequency domain approaches like Bode plots  \cite{book:akt_sysid} for this purpose. 
    \item \textit{During-Estimation}: In this set of approaches, model orders, delay are estimated together with model parameters. Such approaches utilize regularization or applying sparsity constraints on the model parameters while estimating the model parameters to estimate the delay and order \cite{paper:sateesh_CS}. 
    
    \item \textit{Post-Estimation}: In these approaches, a particular model structure is assumed to estimate the model parameters, and the correctness of model order and delay is checked after estimation exercise. Some of the popular approaches are recursive estimation of ARX models, followed by checking under-parameterization or over-parameterization of the model through residual analysis, or hypothesis testing of estimated parameters \cite{book:akt_sysid}. Another widely accepted approach is to estimate a high order model in the first step and subsequently utilize Akaike Information Criteria (AIC) \cite{book:ljung} or frequency-domain based model order reduction \cite{liu2015method}.
\end{enumerate}
This paper aims to develop a theoretically rigorous approach and an automated algorithm that needs minimal user intervention practically for joint estimation of order, delay, and model parameters. For this purpose, we discuss a few relevant modules of matrix factorization or linear algebra methods in the context of system identification. 

It is well known that the Principal Component Analysis (PCA) can be used to compute the solution for the Total Least Squares (TLS) problem \cite{book:joliffe}. The vanilla version of TLS is suited for the case of the homoscedastic errors, meaning the errors in all the variables have the same variance. The generalized case of unequal noise variances in all the variables is dealt with by using weighted TLS.  In  \cite{wentzell1997maximum}, it is shown that the inverse of noise standard deviations should be taken as the optimal weights to obtain maximum likelihood estimates. However, this assumes that the noise standard deviations to be known, which is usually not the case. In \cite{paper:ipca}, an iterative PCA approach is proposed to estimate the unequal noise standard deviations by maximizing the likelihood of model residuals, and simultaneously estimate the model equations and parameters for a steady-state process.

The above methods have been proposed to identify static models. In \cite{paper:dpca}, the PCA approach is extended to determine a difference equation model of a dynamic SISO process. But this approach assumes that both the input and output measurements have the same noise variance (homoscedastic errors). The case of the dynamic models also brings with it the challenge of estimating the input and output orders, delay of the difference equation, and noise variances for heteroscedastic errors. 

These challenges were tackled in  \cite{paper:iecr_dipca}, where the authors propose the Dynamic Iterative PCA (DIPCA) algorithm to estimate all of these parameters along with the model coefficients. This framework was proposed for the output-error model structure, which assumes Gaussian white noise in both the input and output variables. The key idea of the DIPCA algorithm was to scale the measurements with the inverse of a \emph{diagonal} matrix containing the noise standard deviations.

However, this approach cannot be directly used for identifying the ARX model structure, as discussed below briefly and explained comprehensively in Section \ref{sec:ord_noise_known}. The prime reason is the colored nature of output noise, which leads to the fact that the noise covariance matrix is \emph{not diagonal}. The off-diagonal terms in the noise covariance matrix can be parameterized in terms of noise variances and difference equation parameters.   
In this work, we utilize this fact to propose a novel algorithm by appropriately modifying DIPCA  to deal with colored noise in the output, concomitant with the ARX structure.  


The proposed algorithm primarily utilizes the spectral decomposition or singular value decomposition (SVD) of the lagged data matrix to estimate the model parameters. A key contribution of this work is in estimating the error covariance matrix of lagged measurements and using it to appropriately transform the lagged measurements, before applying SVD. This is equivalent to estimating the correct weights for performing the weighted TLS to estimate the ARX model parameters. The error covariance matrix of lagged measurements is obtained through the Yule-Walker equations along with the estimated error variances of white noise sequences $e_y[k]$ and $e_u[k]$ shown in Fig. 1, and estimated model parameters.  A theoretical criterion based on the singular values of the transformed data matrix is used to derive the ARX process order.

The rest of the paper is organized as follows. Section \ref{sec:prelim} defines the identification problem formally, and reviews the literature on identification using the spectral decomposition approach. Section \ref{sec:prop_algo} describes the proposed identification algorithm. Simulation studies presented in section \ref{sec:exp} show the efficacy of the proposed method. Concluding remarks and directions for future work are outlined in section \ref{sec:conc}. 

\section{Preliminaries}
\label{sec:prelim}
We start with the formal description of the full identification problem. A general linear time-invariant (LTI) SISO system model among the noise-free input-output variables can be represented as follows:
\begin{align}
    	y^{\star}[k] + \sum_{i=1}^{n_y}a_{i}y^{\star}[k-i] = \sum_{j=D}^{n_u}b_ju^{\star}[k-j] \label{eq:sisodynamic}
\end{align}
where $D$ is the input-output delay, $n_u$, $n_y$ denotes the input and output orders, respectively.  We also define the process order, $\eta = \max(n_y,n_u)$. We further assume that input measurements are corrupted by Gaussian white noise, while output measurements are corrupted by colored noise following the ARX structure, that is, 

\begin{subequations}
\begin{align}
y[k] &= y^{\star}[k] + v_y[k] \label{eq:out_noise} \\ 
v_y[k] & +\sum_{i=1}^{n_y}a_{i}v_y[k-i] = e_y[k] \\ 
u[k] &= u^{\star}[k] + e_u[k] 
\end{align} \label{eq:noise_add}
\end{subequations}
where $e_y[k] \sim \mathcal{N}(0, \sigma^2_{e_y})$ and $e_u[k] \sim \mathcal{N}(0, \sigma^2_{e_u})$. The noise sequences $e_y[k]$ and $e_u[k]$ are assumed to be independent of each other and also independent of the true values of the inputs and outputs. The complete identification problem is to estimate the following parameters from $N$ samples of noisy measurements:
\begin{enumerate}
    \item The input-output delay (D).
    \item The input-output orders ($n_u,n_y$). 
    \item The noise variances, $\sigma^2_{e_y}$ and $\sigma^2_{e_u}$. 
    \item The coefficients, $\{a_{i}\}_{i=1}^{n_y}$ and $\{b_{j}\}_{j=D}^{n_u}$. 
\end{enumerate}
It may be noted that the autocovariance function (ACVF) of output noise can be parameterized using the noise variance  $\sigma^2_{e_y}$ and the model parameters $\{a_{i}\}_{i=1}^{n_y}$. This is due to the particular structure of ARX noise. We denote the variance of output error $v_y[k]$ by $\sigma^2_{v_y}$.

We start with a brief review of the DIPCA algorithm \cite{paper:iecr_dipca}. This algorithm was proposed for a different model structure and provide consistent estimates under different assumptions, as discussed later. Our focus is to choose the critical relevant ideas from this work and modify them suitably for EIV-ARX model identification. 
\subsection{Dynamic Iterative PCA}
DIPCA \cite{paper:iecr_dipca}  was proposed to deal with errors in input and output measurements with unequal and unknown error variances.  The method estimates the noise variances and uses its inverse square-root to scale the measurements, which basically transforms the heteroscedastic error case to the homoscedastic case.  A fundamental assumption made in the method is that the errors in input and output measurements are white and also mutually independent, which is not valid for the EIV-ARX model considered in this work.  Nevertheless, we wish to examine the model's quality obtained by applying DIPCA for the EIV-ARX case.  We describe in brief the essential ideas involved in this approach and refer the reader to \cite{paper:iecr_dipca} for a detailed description.

It should be noted that the LTI difference equation in \eqref{eq:sisodynamic} can be expressed as the following constraint:
\begin{align}
    \mathbf{\theta}^T \mathbf{z}_{\eta}^{\star}[k] = 0 
    \label{eq:nullsp}
\end{align}
\vspace{-0.7 cm}
{\footnotesize \begin{subequations}
		\begin{align*}
		\mathbf{\theta} &=  \begin{bmatrix}
		1 \  a_1 \ \ldots \ a_{n_y} \   \mathbf{0}_{\eta-n_y+D} \  -b_D \ \ldots \ -b_{n_u} \  \mathbf{0}_{\eta-n_u} 
		\end{bmatrix}^T\\ 
		\mathbf{z}^{\star}_{\eta}[k]  &= \begin{bmatrix}
		y^{\star}[k] \  \ldots \  y^{\star}[k - \eta] \  u^{\star}[k] \  \ldots \  u^{\star}[k - \eta]
		\end{bmatrix}^T
		\end{align*} 
\end{subequations}} %
We refer to $\mathbf{\theta}$ as the parameter vector and $\mathbf{z}_{\eta}^{\star}[k]$ as a vector of appropriately lagged noise-free input-output variables. In this case, the identification problem is to estimate the parameter vector of the difference equation from $N-\eta$ samples of $\mathbf{z}_{\eta}^{\star}[k]$, denoted by $\mathbf{Z}^{\star}_{\eta}$:
\begin{align}
    \mathbf{Z}^{\star}_{\eta} = \begin{bmatrix}
	\mathbf{z}^{\star}_{\eta}[\eta+1] &  \mathbf{z}^{\star}_{\eta}[\eta + 2] & \ldots & \mathbf{z}^{\star}_{\eta}[N]
	\end{bmatrix}^T \label{eq:nf_stack}
\end{align}
The constraint model in Eq. \eqref{eq:nullsp} can also be inferred as the parameter vector ($\mathbf{\theta}$) being orthogonal to $\mathbf{z}^{\star}[k]$ for all $k = \{\eta+1,\ldots,N\}$ and the same can be expressed as:
\begin{align*}
    \mathbf{Z}^{\star}_{\eta} \mathbf{\theta} = \mathbf{0}
\end{align*}
If the system order is known, then the SVD of the matrix $\mathbf{Z}^{\star}_{\eta}$ or equivalently PCA can be used for the estimating parameter vector $\mathbf{\theta}$. We use the analogous approach of eigenvalue decomposition (EVD) of the sample covariance matrix $\mathbf{S}_{Z^{\star}_{\eta}} = \frac{1}{N} \mathbf{Z^{\star}_{\eta}}^T\mathbf{Z^{\star}_{\eta}}$ for deriving the null space of $\mathbf{Z^{\star}_{\eta}}$:
\begin{align}
    \mathbf{S}_{Z^{\star}_{\eta}}\mathbf{V^{\star}_{0}} = \mathbf{V^{\star}_{0}} \mathbf{\Lambda^{\star}_{0}}
\end{align}
where $\mathbf{\Lambda^{\star}_{0}}$ is a diagonal matrix consisting of the eigenvalues (ordered from largest to smallest), and the columns of $\mathbf{V^{\star}_{0}}$ contains the corresponding eigenvectors. The eigenvector corresponding to zero eigenvalue (smallest eigenvalue) can be used to recover the parameter vector ($\mathbf{\theta}$) \cite{paper:rao1964pca,book:joliffe,paper:ipca}. 

We proceed to the case of noisy measurements. The noisy measurements are generated by corrupting the noise-free output measurements with the colored noise following ARX structure, and the input is corrupted with white noise, as shown in Eq. \eqref{eq:noise_add}. We generate $N$ samples of noisy input-output measurements and construct the lagged data matrix of measurements as follows:

{\footnotesize
\begin{subequations}
\begin{align}
    \mathbf{z}_{\eta}[k]  & = \begin{bmatrix}
		y[k]  & \ldots & y[k - \eta] & u[k-D]  \ldots u[k-\eta]\end{bmatrix} \\
	\mathbf{Z}_{\eta}  &= \begin{bmatrix}
	\mathbf{z}_{\eta}[\eta + 1] &  \mathbf{z}_{\eta}[\eta + 2] & \ldots & \mathbf{z}_{\eta}[N]	\end{bmatrix}^T 
\end{align} 
\label{eq:stack_perf}
\end{subequations}}

The key idea used in DIPCA is to scale the measurements using the inverse-square root of the error covariance matrix as follows:
\begin{align}
    \mathbf{z}_{\eta,s}[k] = \mathbf{\Sigma}^{-1/2}_{e_{\eta}}\mathbf{z}_{\eta}[k] \label{eq:scal_vec}
\end{align}
where $\mathbf{z}_{\eta,s}[k]$ is the scaled version of noisy measurements and $\mathbf{\Sigma}_{e_{\eta}}$ is a diagonal noise covariance matrix defined as
\begin{align}
    \mathbf{\Sigma_{e_{\eta}}} = \begin{bmatrix} \sigma^2_{v_y}\mathbf{I}_{\eta + 1}	&  \mathbf{0} \\ 
	\mathbf{0} & \mathbf{ \sigma^2}_{e_u}\mathbf{I}_{\eta+1}
	\end{bmatrix} \label{eq:cov_Leta}
\end{align}
where $\mathbf{I}_{\eta + 1}$ denotes a identity matrix of dimension $\eta + 1$.

PCA is applied to the scaled data matrix $\mathbf{Z}_{\eta,s}$, and the eigenvectors corresponding to the smallest singular value is used to obtain the model parameter vector estimate $\mathbf{\hat{\theta}}_s$ which relates the scaled lagged variables.  The estimate of the model parameters in terms of the original variables is obtained as 
\begin{align} \label{eq:modelparamscaled}
 \mathbf{\hat{\theta}} = \mathbf{\hat{\theta}}_s\mathbf{\Sigma}^{-1/2}_{e_{\eta}}
\end{align}
However, in order to use the above approach, estimates of the error variances and the model order are required.  These are estimated in DIPCA by first constructing a lagged data matrix $\mathbf{Z}_{L}$ using a sufficiently large lag $L$ greater than $\eta$ and applying the IPCA algorithm \cite{paper:ipca} to this lagged data matrix.  Due to the use of a large lag, multiple constraints relating the lagged variables are obtained.  The number of such constraints $d$ can be estimated by examining the number of unity singular values of the scaled data matrix $\mathbf{Z}_{L,s}$ (where subscript $L$ and $s$ denote the lag order and scaling of measurements).  Moreover, it can be shown that the number of constraints and model order satisfies the following equation, 
\begin{align}
    d = L - \eta + 1 \label{eq:ordest}
\end{align}
which allows the model order to be estimated.  The error variances are also estimated using an iterative scheme in DIPCA.

We use an illustrative case-study to demonstrate the use of EVD in identification as this is a vital concept in this work. Consider a second-order, unit delay LTI system:
\begin{align}
    y^{\star}[k] - 1.5 y^{\star}[k-1] + 0.7 y^{\star}[k-2] =  u^{\star}[k-1] + 0.5u^{\star}[k-2] \label{eq:secondord_sys}
\end{align}
The input $u^{\star}$ is chosen as a full length and full band pseudo-random binary signal (PRBS) of length $4095$. The output, $y^{\star}$ is generated according to Eq. \eqref{eq:secondord_sys}. Noisy measurements are generated using:
\begin{subequations}
\begin{align}
v_y[k] &- 1.5 v_y[k-1] + 0.7 v_y[k-2] = e_y[k] \\
y[k] &= y^{\star}[k] + v_y[k] \\
u[k] &= u^{\star}[k] + e_u[k]  
\end{align} \label{eq:ex_noise1}
\end{subequations}
where $e_y[k] \sim \mathcal{N}(0, \sigma^2_{e_y})$, $e_u[k] \sim \mathcal{N}(0, \sigma^2_{e_u})$ and $\sigma^2_{e_y}$, $\sigma^2_{e_u}$ are chosen to be 0.2 and 0.1 respectively to maintain signal to noise ratio of 10. The task is to estimate the coefficients of the difference equation from the noisy measurements. 

The DIPCA algorithm is applied to the second order system defined in Eq. \eqref{eq:secondord_sys}, ignoring the fact that the noise corrupting the output measurements is coloured. 
The results from 100 runs of MC simulations are presented in second column of Table \ref{tab:runex_all}.  

\begin{table}[H]
	\caption{MC simulation results for various algorithms}
	\label{tab:runex_all}
	\centering
	\begin{footnotesize}
		\begin{tabular}{cc|c|c|c|c}
			\hline
			{} & {} &  \multicolumn{2}{|c|}{DPCA} & \multicolumn{2}{|c}{DIPCA} \\
			\hline
			  &Value& $\mu $ & $2\sigma$& $\mu $ & $2\sigma$  \\\hline
			$a_1$ & $-1.5$ & -1.552 &   0.0186   & -1.498  &  0.026 \\ 
			$a_2$ & $0.7$ & 0.748  &  0.0202& 0.699 &    0.025  \\
			$b_0$ & $0$ & 0.003 &    0.0340   & 0.002 &   0.046 \\ 
			$b_1$ & $1$ & 0.967 &   0.0279   & 1.102  &  0.034 \\ 
			$b_2$ & $0.5$ &  0.432 &   0.0378   & 0.559 &   0.045 \\ \hline 
		\end{tabular}
	\end{footnotesize} 
\end{table}

For the purpose of comparison, we also apply dynamic PCA algorithm \cite{paper:dpca} which is a special case of homoskedastic errors in DIPCA algorithm \cite{paper:iecr_dipca}. The parameter vector is estimated from the eigenvector corresponding to smallest eigenvalue of the (unscaled) sample covariance matrix, defined as $\mathbf{S}_{Z_{\eta}} = \frac{1}{N} \mathbf{Z}_{\eta}^T \mathbf{Z}_{\eta}$. We perform 100 runs of Monte-Carlo simulations and present the numerical details in the first column of Table \ref{tab:runex_all}. 

It can be observed from the results presented in the first column of Table \ref{tab:runex_all} that the estimates obtained using DPCA are biased. It can also be observed that the estimates obtained using DIPCA are more accurate as compared to DPCA. However, a biased estimates of parameters $b_1$ and $b_2$ are obtained. This is expected as the fundamental idea of DIPCA \cite{paper:iecr_dipca} algorithm is to scale the measurements with the inverse square root of the noise covariance matrix, which is assumed to be diagonal. This assumption is valid for the case of measurements corrupted with Gaussian white noise. For an EIV-ARX process, the output measurements are corrupted with colored noise, and hence a diagonal noise covariance matrix is not the optimal choice. 

In the next section, we propose a novel algorithm that addresses the issue of colored noise using a spectral decomposition framework by appropriately choosing the elements of the noise-covariance matrix. 

\section{Proposed Algorithm}
\label{sec:prop_algo}
In the previous section, we demonstrated that existing methods like DPCA and DIPCA produce biased estimates, fundamentally because they do not correctly account for the noise characteristics that corrupt the measurements. In this work, we propose a modification of the DIPCA algorithm to deal with the noise characteristics of an EIV-ARX model.  The additional challenges considered in this work are estimating the noise variances, delay, input-output orders, and ARX model parameters. 

For ease of understanding, we present the proposed algorithm in three subsections.  In the first sub-section, we assume that process order and variances of noise sources are known and describe how the ARX model coefficients can be estimated. In the next sub-section, we assume that the process order is unknown, but noise variances are known, and outline an approach for estimating the process order.  Finally, in the last sub-section, we treat the problem of estimating all model parameters, including the noise variances. 

\subsection{Model identification for known order \& noise variances} 
\label{sec:ord_noise_known}
We first describe how the model coefficients of an EIV-ARX model can be iteratively estimated when the system order ($\eta$) and noise variances ($\sigma^2_{e_y}$ and $\sigma^2_{e_u}$) are known.  It should be noted that the noise in the columns of $\mathbf{Z}_{\eta}$ defined in Eq. \eqref{eq:stack_perf} are correlated due to the colored nature of output noise. For the ARX model structure, the noise covariance matrix of lagged measurements is given by:

\begin{align}
    \mathbf{\Sigma_{e_{\eta}}} = \begin{bmatrix}
\mathbf{ \Sigma_{v_y}}	&  \mathbf{0}\\ 
	\mathbf{0} & \mathbf{ \sigma^2}_{e_u}\mathbf{I}_{\eta+1}
	\end{bmatrix} \label{eq:cov_L}
\end{align}
where $\mathbf{ \Sigma_{v_y}}$ is the output noise covariance matrix given by
{\footnotesize 
\begin{align}
\mathbf{ \Sigma_{v_y}}=	\begin{bmatrix}
	\sigma_{vv}[0] & \sigma_{vv}[1] & \sigma_{vv}[2]  & \ldots &\sigma_{vv}[\eta] \\
	\sigma_{vv}[1] & \sigma_{vv}[0] & \sigma_{vv}[1]  & \ddots & \sigma_{vv}[\eta-1] \\
	\sigma_{vv}[2] & \sigma_{vv}[1] & \sigma_{vv}[0] & \ddots & \vdots \\
	\vdots &  \ddots & \ddots &  \ddots  & \vdots  \\
	\sigma_{vv}[\eta] & \sigma_{vv}[\eta-1] & \ldots & \ldots & \sigma_{vv}[0]
	\end{bmatrix}  \label{eq:cov_outL}
\end{align}}
where $\sigma_{vv}[l]$ denotes the auto-covariance function (ACVF) of sequence $v_y[k]$ at lag $l$. 

We transform the measurements using the inverse square root of $\mathbf{\Sigma_{e_{\eta}}}$ as shown in Eq. \eqref{eq:scal_vec} and perform the EVD of the covariance matrix for the transformed measurements: 
\begin{align}
    \mathbf{S}_{Z_{\eta,s}} = \frac{1}{N} \mathbf{\Sigma}^{-1/2}_{e_{\eta}} \mathbf{Z}_{\eta}^T \mathbf{Z}_{\eta} \mathbf{\Sigma}^{-1/2}_{e_{\eta}} \label{eq:cov_scalZ}
\end{align}
We can show that
\begin{align}
\mathbf{\Sigma}_{Z_{\eta,s}} = E[\mathbf{S}_{Z_{\eta,s}}] =  \mathbf{S}_{Z^{\star}_{\eta,s}} + \mathbf{I}_{\eta+1} \label{eq:eig_scal}
\end{align}
where $\mathbf{S}_{Z^{\star}_s}$ is the covariance matrix for scaled noiseless measurements, defined as $\mathbf{S}_{Z^{\star}_{\eta,s}} = \frac{1}{N} \mathbf{\Sigma}_e^{-1/2} \mathbf{Z^{\star}_{\eta}}^T \mathbf{Z^{\star}_{\eta}} \mathbf{\Sigma}_e^{-1/2}$. Two important conclusions can be inferred from Eq. \eqref{eq:eig_scal}. We can easily verify that the eigenvectors of $\mathbf{\Sigma}_{Z_{\eta,s}}$ and $\mathbf{S}_{Z^{\star}_{\eta,s}}$ are identical.   Thus, the model parameters can be estimated using the eigenvector of $\mathbf{\Sigma}_{Z_{\eta,s}}$ corresponding to its smallest eigenvalue, after re-transforming in terms of original variables, similar to Eq. \eqref{eq:modelparamscaled}.
We can also verify that the eigenvalues of $\mathbf{\Sigma}_{Z_{\eta,s}}$ are equal to the eigenvalues of $\mathbf{S}_{Z_{\eta,s}}$ increased by unity.  Thus, the smallest eigenvalue of $\mathbf{\Sigma}_{Z_{\eta,s}}$ will be equal to unity.  In fact, these properties hold for any lag used to construct the lagged measurements. 
In practice, the sample covariance matrix  $\mathbf{S}_{Z_{\eta,s}}$ is used instead of $\mathbf{\Sigma}_{Z_{\eta,s}}$, since it is a consistent estimate.

In order to apply the above procedure, the noise covariance matrix of lagged measurements is required.  The output noise covariance matrix defined in Eq. \eqref{eq:cov_outL} is a function of the variance $\sigma^2_{e_y}$ and coefficients $\{a_{i}\}_{i=1}^{n_y}$. 

We use Yule-Walker equations for the estimation of auto-covariance function of the output noise shown in Eq. \eqref{eq:out_noise} which can be expressed as:
\begin{align}
    \sum_{k=0}^{n_y} a_k \sigma_{vv}[l-k]=\begin{cases}\sigma_{e_y}^2 \;, \quad l=0\\ 0 \;\; , \quad l>0  \end{cases} \label{eq:yule_walker}
\end{align}
where $a_0 = 1$ and $l$ is a positive integer. It can be seen that these are a set of $n_y+1$ linear equations in $n_y+1$ variables for $l = \{0,1,\ldots,n_y\}$. Hence the solution of $\sigma_{vv}[l]$ for $l = \{0,1,\ldots,n_y\}$ can be easily determined. It should be noted that ACVF is symmetric, that is $\sigma_{vv}[-l] = \sigma_{vv}[l]$, where $l$ can be any integer. The ACVF for $l>n_y$ can also be obtained using Eq \eqref{eq:yule_walker} recursively. We use the estimated ACVF to propose the following iterative procedure for the estimation of model parameters. 

We start with a guess of $\mathbf{\Sigma_{e_{\eta}}}$ as identity matrix and apply PCA to obtain an initial estimate of the model parameters.  We use the estimated model parameters to construct $\mathbf{\Sigma_{v_y}}$ using Eq. \eqref{eq:yule_walker} and Eq. \eqref{eq:cov_outL}, which is further used to construct $\mathbf{\Sigma_{e_{\eta}}}$ as defined in Eq. \eqref{eq:cov_L}.  The updated estimate of $\mathbf{\Sigma_{e_{\eta}}}$ can now be used to transform the lagged measurements as in Eq. \eqref{eq:cov_scalZ}, and new estimates of model parameters are obtained.  This iterative procedure is repeated until the convergence of model parameters. 

We apply this approach for the system presented in Eq. \eqref{eq:secondord_sys}. The average of the smallest eigenvalue from 100 runs of MC simulations is observed to be $1.002$ for the transformed measurements. The estimates of the model coefficients obtained from the eigenvector corresponding to smallest eigenvalue are reported in the third column of Table \ref{tab:runex_all_1}. 

\begin{table}[H]
	\caption{MC simulation results for various algorithms}
	\label{tab:runex_all_1}
	\centering
	\begin{footnotesize}
		\begin{tabular}{cc|c|c|c|c}
			\hline
			{DIPCA} & {} &  \multicolumn{2}{|c|}{Unknown diagonal $\mathbf{\Sigma_e}$} & \multicolumn{2}{|c}{ Known non-diagonal  $\mathbf{\Sigma_e}$}\\
			\hline
			  &Value&  $\mu $ & $2\sigma$  & $\mu $ & $2\sigma$ \\\hline
			$a_1$ & $-1.5$   & -1.498  &  0.026  & -1.5 & 0.024 \\ 
			$a_2$ & $0.7$ & 0.699 &    0.025 & 0.7 & 0.025 \\
			$b_0$ & $0$ & 0.002 &   0.046 & 0.002 & 0.036\\ 
			$b_1$ & $1$ & 1.102  &  0.034 & 1.003  & 0.037 \\ 
			$b_2$ & $0.5$ & 0.559 &   0.045 & 0.503 & 0.042 \\ \hline 
		\end{tabular}
	\end{footnotesize} 
\end{table}
It can be observed from the last two columns of Table \ref{tab:runex_all_1} that unbiased estimates of all model parameters are obtained if the correct (non-diagonal) structure of the noise covariance matrix is utilized. In this subsection, we used the correct structure of the noise covariance matrix to obtain unbiased estimates of the parameters, when system order and error variances are known. In the next two subsections, this idea is extended to estimate both the system order and model parameters.

\subsection{Model Identification for known noise variances ($\sigma^2_{e_y}$ \& $\sigma^2_{e_u}$)} 
\label{sec:noise_known}
In this subsection, we consider the noise variances to be known, but the system order to be unknown. If we can estimate the system order, we can utilize the approach described in the preceding subsection to estimate the model parameters. 

Since the system order is unknown, we construct a lagged data matrix using a sufficiently large lag of $L$ greater than the system order.  The lagged data samples and the corresponding data matrix are defined as follows: 
\begin{align}
		\mathbf{z}_{L}[k] & = \begin{bmatrix} y[k] &  \cdots & y[k-L]   & u[k] \cdots & u[k-L]
		\end{bmatrix}^T  \nonumber \\ 
		\mathbf{Z}_L & = \begin{bmatrix} \mathbf{z}_{L}[L+1] & \mathbf{z}_{L}[L+2] & \cdots & \mathbf{z}_{L}[N]
		\end{bmatrix}^T  \label{eq:Z_data_gen_L} 
\end{align}

The use of a lag $L$ greater than the system order $\eta$ results in multiple linear relations relating lagged input-output variables. Each relation corresponds to a shifted version of the discrete dynamical model given by Eq. (\ref{eq:sisodynamic}).  It can be shown that the number of independent linear relations or constraints $d$ that relates the lagged variables is given by Eq. (\ref{eq:ordest}).  Furthermore, using Eq. (\ref{eq:eig_scal}), which is valid for any lag $L$, we can show that the number of unity eigenvalues of $\mathbf{S}_{Z_{L,s}}$ (in the limit as the number of samples tend to infinity) corresponds to the number of linear relations among the lagged variables.  Thus, it is easy to estimate $d$ by examining the \textit{unity} eigenvalues of the scaled data covariance matrix and hence the system order can be estimated as
\begin{align}
\hat{\eta} = L - d + 1
\end{align}

Although the system order can be correctly estimated, the model parameters cannot be directly estimated using the eigenvector corresponding to the smallest eigenvalue, as discussed in the preceding section. This is because the multiple linear equations obtained from eigenvectors corresponding to unity eigenvalues can be linear combinations of the discrete shifted model equations \cite{paper:iecr_dipca, mann2020optimal}. A procedure to recover the model parameters from these eigenvectors is described.

The constraints relating the noise-free variables corresponding to a stacked vector $\mathbf{z}_L$ can be written as

\begin{align}
    \begin{bmatrix}
    \mathbf{A} & -\mathbf{B}
    \end{bmatrix} & \begin{bmatrix} \mathbf{y}^{\star}_L[k] \\  \mathbf{u}^{\star}_L[k] \end{bmatrix} = \mathbf{0} \label{eq:resi_L3}
\end{align}
where
{\footnotesize
\begin{align} 
\mathbf{A} &= \begin{bmatrix}
       1 & a_1 & \ldots &  a_{n_y} &  0 & 0 & \ldots & 0 & \ldots & 0 \\
       0 & 1 & a_1 & \ldots & a_{n_y} & 0 & \ldots & 0 & \ldots & 0 \\
       \vdots & \vdots & \vdots & \vdots & \vdots \\
       0 & 0 & 0 & 0 & \ldots & 0 & 1 & a_1 & \ldots & a_{n_y}
    \end{bmatrix}
\end{align}
\begin{align}
    \mathbf{B} = \begin{bmatrix}
        \mathbf{0}_D & b_D & \ldots & b_{n_u} & 0 & \ldots & \ldots & \ldots & 0 \\
        \mathbf{0}_{D} & 0 &  b_D & \ldots & b_{n_u} & 0 & \ldots & \ldots & 0  \\
        \vdots & \vdots & \vdots & \vdots & \vdots & \vdots & \vdots & \vdots \\
        \mathbf{0}_{D} & 0 & 0 & \ldots & 0 & 0 & b_{D} & \ldots & b_{n_u}     \end{bmatrix} \label{eq:resi_L3_a}
\end{align}}
Let $\mathbf{V}_2^T$ be the $d$ eigenvectors of scaled covariance matrix $\mathbf{S}_{Z_{L,s}}$ corresponding to the smallest $d$ eigenvalues (all of which will be approximately equal to unity). An estimate of the constraint matrix can be obtained by suitable rotation of these eigenvectors given by 
\begin{align} \label{eq:rotationmatrix}
        \begin{bmatrix}
    \mathbf{\hat{A}} & -\mathbf{\hat{B}}     \end{bmatrix} = \mathbf{R}\mathbf{V^T_2}\mathbf{\Sigma}^{-1/2}_{e_L}
\end{align}
where $\mathbf{R}$ is a $d \times d$ rotation matrix.  The rotation matrix can be computed by making use of the known structure of $\mathbf{A}$. It is noted that each row of $\mathbf{A}$ is just a shifted version of the previous row. Although the non-zero elements of the matrix $\mathbf{A}$ are unknown, the locations of ones and zeros in the matrix $\mathbf{A}$ are known based on the system order.  The rotation matrix is chosen such that the estimated matrix $\mathbf{\hat{A}}$ in Eq. (\ref{eq:rotationmatrix}) has the same structure as $\mathbf{A}$ with respect to the unity and zero elements.

The rotation matrix  $\mathbf{R}$ has $d^2$ unknowns, and hence we require at least $d^2$ equations to arrive at a unique solution of $\mathbf{R}$. Each row of $\mathbf{A}$ contains $L-\hat{\eta}$ zeros due to excessive stacking and hence the total number of zeros in $d$ rows of $\mathbf{A}$ is $d(L-\hat{\eta}) = d(d-1)$. Each row of $\mathbf{A}$ contains $1$ along the diagonal entries. Hence the total number of known elements of $\mathbf{A}$ is equal to $d(d-1) + d = d^2$. This shows that the rotation matrix can be derived using an exactly determined system of linear equations. From the rows of the rotated matrix, the model parameters can be obtained as the average of the appropriate elements. A similar approach is used in \cite{paper:dycops_dipca} for the estimation of rotation matrix. 
Using the estimated model parameters, the error covariance matrix of lagged measurements can be computed to re-scale the lagged data.  A new estimate of the model parameters can be obtained, and the procedure is repeated until convergence.
\subsection{Estimation of error covariance matrix}
\label{sec:noise_covest}
In this subsection, we assume the error variances to be unknown and describe a procedure to estimate them by maximizing the likelihood of residuals without using any information about the delay or order of the system. 

It should be noted that there are only two parameters, $\sigma^2_{e_y}$ and $\sigma^2_{e_u}$ that have to be estimated for recovering the entire noise covariance matrix described in Eq. (\ref{eq:cov_L}).  
The noise covariance matrix, $\mathbf{ \Sigma_{v_y}}$ in Eq. \eqref{eq:cov_outL} can be obtained using the coefficients $\{a_i\}^{n_y}_{i=1}$ because the output noise follows ARX structure as shown in Eq. \eqref{eq:noise_add}. 

We recall from the preceding section that by using a lag $L$ greater than $\eta$ results in multiple independent linear relations among the lagged variables.  An estimate of the constraint matrix can be obtained from the eigenvectors corresponding to unity eigenvalues of the scaled data matrix.    

We use these constraints for the estimation of $\sigma^2_{e_y}$ and $\sigma^2_{e_u}$ by defining the residuals $\mathbf{r}_L[k]$ as:
\begin{align}
        \mathbf{r}_L[k] = \begin{bmatrix}
    \mathbf{\hat{A}} & -\mathbf{\hat{B}}
    \end{bmatrix}  \mathbf{z}_L[k]
\end{align}
It can be proved that the constraint residuals follow a normal distribution with zero mean and covariance matrix defined by 
\begin{align} \mathbf{\Sigma_{r_L}} =  \begin{bmatrix}
    \mathbf{\hat{A}} &  -\mathbf{\hat{B}}
    \end{bmatrix} \mathbf{\Sigma_{e_{L}}} \begin{bmatrix}
       \mathbf{\hat{A}} &  -\mathbf{\hat{B}}
    \end{bmatrix}^T 
\end{align}
An estimate of  $\sigma^2_{e_y}$ and $\sigma^2_{e_u}$ can be obtained by maximizing the joint density function of $\mathbf{r}_L[L+1],\mathbf{r}_L[L+2],\ldots, \mathbf{r}_L[N]$.  This results in the following optimization problem.
\begin{align}
\min_{\sigma^2_{e_y},\sigma_u^2}& N \log \vert \mathbf{\Sigma_{r_L}} \vert
 + \sum_{k=L+1}^{N} \mathbf{r}_L^{T}[k] \mathbf{\Sigma_{r_L}^{-1}}\mathbf{r}_L[k] \label{eq:Sigmae_est2}
\end{align}
Using the estimated values of the noise variances, the error covariance matrix of lagged measurements can be computed using Eq. \eqref{eq:yule_walker}, as described in section \ref{sec:noise_known}. 

It may be noted that although the proposed algorithm provides estimates of all unknown parameters, including model parameters, it is possible to perform one final refinement to estimate the model parameters. The estimated \emph{model order and error variances} obtained from this algorithm can be used to construct the lagged data matrix using a lag $L$ equal to $\hat{\eta}$.  The method described in section \ref{sec:ord_noise_known} can now be used to obtain the model parameter estimates from the eigenvector corresponding to the smallest (unity) eigenvalue.  This refinement is adopted in this paper.

The proposed algorithm is summarized in Table \ref{tab:prop_algo}. In the next section, we assess the performance of the proposed algorithm on simulated examples.

\begin{table}[H]
\caption{Proposed Algorithm for ARX-EIV Models \label{tab:prop_algo}}
\hrulefill
\begin{enumerate}
    \item Stack the measurements up to a sufficiently high lag $L$ as shown in Eq. \eqref{eq:Z_data_gen_L}. Kick-start the algorithm with a guess for number of constraints $d_{\text{guess}}$ = $L$. Let $\{\mathbf{\Sigma_{e_L}}\}^{i=0} = \mathbf{I}$, where $i$ is the iteration counter. 
    \item Estimate the constraint matrix $\{\mathbf{A}\}^{i+1}$ for given noise covariance matrix $\{\mathbf{\Sigma_{e_L}}\}^{i}$ as discussed in section \ref{sec:noise_known}. 
    \item Estimate the noise covariance matrix $\{\mathbf{\Sigma_{e_L}}\}^{i+1}$ for given constraint matrix $\{\mathbf{A}\}^{i+1}$ as discussed in section \ref{sec:noise_covest}.
    \item Repeat steps 2 \& 3 until the convergence of noise covariance matrix or eigenvalues of covariance matrix. 
    \item Recover the number of linear relations ($\hat{d}$) from the cardinality of unity eigenvalues. If $\hat{d} \neq d_{\text{guess}}$, decrease $d_{\text{guess}}$ and rerun steps 2-4. 
    \item If $\hat{d} = d_{\text{guess}}$, estimate coefficients using converged noise variances and order obtained in steps 4 \& 5 as shown in section \ref{sec:ord_noise_known}.  
\end{enumerate}
\hrulefill
\end{table}
\section{Experiments}
Two case studies in this section demonstrate the functioning and efficacy of the proposed algorithm. Both the case-studies show the working of the proposed algorithm. The first case study is focused on explaining the behavior of the proposed algorithm in terms of the quality of estimated parameters as a function of stacking order and sample size. 

The second example presents a comparative study of estimated parameters by the proposed algorithm and the prediction error method (PEM) developed for identification of non-EIV ARX models at low and high SNR for input measurements. If the SNR for input measurements tends to infinity, then the PEM method will provide consistent estimates, and we can assess whether our proposed method can also provide equally good estimates in this case. 

\label{sec:exp}
\subsection{Example 1: SISO Second order system}
\label{ex:ex1}

\begin{figure}[h]
      \centering
      \includegraphics[scale=0.25]{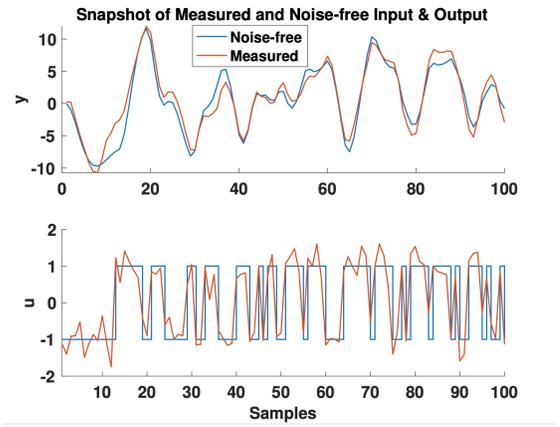}
      \caption{A snapshot for the input and output data in Example \ref{ex:ex1}}
      \label{fig:iodata_ex1}
\end{figure}
\begin{figure}[H]
      \centering
      \includegraphics[scale=0.17]{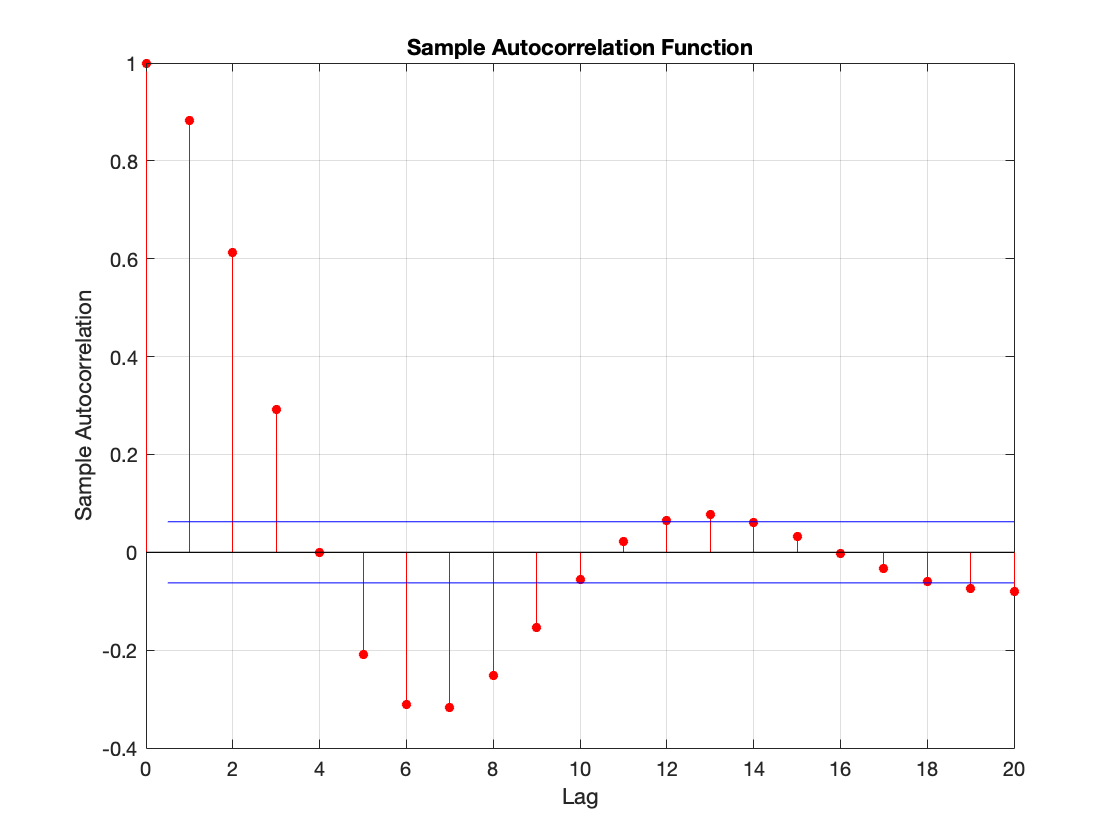}
      \caption{ACF of output noise in Example \ref{ex:ex1}}
      \label{fig:ex1_outnoise}
\end{figure}

Consider the same second-order system described in Eq. \eqref{eq:secondord_sys}. The system is simulated with PRBS input of length $1023$ to generate noise-free measurements. The output is further corrupted with noise following ARX structure, and the input is corrupted with white noise. The variance values for $\sigma^2_{e_y}$ and $\sigma^2_{e_u}$ are  chosen as $0.4$ and $0.1$ to maintain signal to noise ratio (SNR) of 10. We define the SNR of output and input as the variance of the noise-free signal and the corresponding noise, as shown below: 
\begin{align*}
    \text{SNR}_{\text{output}} = \frac{\text{var}(y^{\star})}{\text{var}(v_y)}, \qquad 
    \text{SNR}_{\text{input}} = \frac{\text{var}(u^{\star})}{\text{var}(e_u)} 
\end{align*}
Thus we obtain 1023 samples of noisy input and output measurements which are used for the identification. A snapshot of few samples is shown in Figure \ref{fig:iodata_ex1}. 
The autocorrelation function and partial autocorrelation function of the noise in output measurement is also shown in Figure \ref{fig:ex1_outnoise}. These plots signify the ARX nature of the noise.

The proposed algorithm is applied under the assumption that no prior information is known about the system, and hence we stack the lagged variables with suitably large stacking order $L = 5$ to construct $\mathbf{Z}_L$ as defined in Eq. \eqref{eq:Z_data_gen_L}. As the order is unknown, we execute the proposed algorithm starting with $d_{\text{guess}} = 5$ and examine the converged eigenvalues of the scaled data matrix for each guess. The covariance matrix of  $\mathbf{Z}_{s,L}$ will have $2(L+1) = 12$ eigenvalues but we present the last six eigenvalues in Table \ref{tab:ex1_eig} due to our focus on the small eigenvalues. 

\begin{table}[H]
\caption{Eigenvalues for Example 1}
\label{tab:ex1_eig}
\begin{center}
\begin{tabular}{c|c}
\toprule
$d_{\text{guess}}$   & Eigenvalues (last 6) \\
\midrule 
5  & $\begin{bmatrix} 2.8 &   1.2 &    0.46 &    0.45 &   0.43 &  0.42 \end{bmatrix}$ \\
4  & $\begin{bmatrix} 11.2 & 5.7 & 1.03 &1.01 & 0.97 & 0.97 \end{bmatrix}$ \\
\bottomrule
\end{tabular}
\end{center}
\end{table}
From Table \ref{tab:ex1_eig}, It can be clearly observed that the last $5$ eigenvalues are not equal to unity for $d_{\text{guess}} = 5$, while for $d_{\text{guess}} = 4$, it can be observed that the last $4$ eigenvalues are approximately unity. This can also be confirmed from hypothesis testing for equality of the smallest $d_{guess}$ eigenvalues, which are expected to be unity \cite{mann2020optimal}. The results for hypothesis testing for each value of $d_{\text{guess}}$ are reported in Table \ref{tab:Ex1hypotest}. It can be observed that the null hypothesis is rejected for $d_{\text{guess}} = 5$ and not rejected for $d_{\text{guess}} = 4$. Thus, the results clearly indicate that there are $4$ linear relations among the $12$ lagged input-output variables. The order of the ARX model can be computed as $\hat{\eta} = L - \hat{d} + 1 = 5- 4 + 1 = 2$, which corresponds to the true process order. 
\begin{table}[H]
\caption{Hypothesis test results for eigenvalues in Example \ref{ex:ex1}}
    \label{tab:Ex1hypotest}
    \begin{center}
    \begin{tabular}{c|c|c|c} \toprule
    $d_{\text{guess}}$ & Degrees of freedom ($\nu$) & Test statistic &  Test criterion ($\alpha$ = 0.05) \\ \midrule
    5 & 14 & 1659 & 21 \\
4 & 9 & 1.49 & 14.68  \\ 
\bottomrule
    \end{tabular}
    \end{center}
\end{table}
The true and estimated auto-covariance function of the output noise for the correctly estimated system order of $2$ is presented in Table \ref{tab:ex1_acvf}. It can be observed that the estimated auto-covariance function is close to the true value used for simulating the data. The input noise variance is estimated as $0.09$, which is also close to the true value of $0.1$ used in the simulation. The variance of the excitation signal of output noise (denoted by $\sigma^2_{e_y}$) is estimated to be $0.21$, which is close to the true value of $0.2$.
\begin{table}[H]
\caption{ACVF of output noise in Example 1}
\label{tab:ex1_acvf}
\begin{center}
\begin{tabular}{c|c|c|c|c|c|c}
\toprule
Lag & 0 & 1 & 2 & 3 & 4 & 5\\
\midrule 
True    &   1.77 &   1.56 &    1.1 &   0.56 &     0.07  &   -0.28  \\
\hline 
Estimated  & 1.7 & 1.5 & 1 & 0.48 & 0.015 & -0.31  \\
\bottomrule
\end{tabular}
\end{center}
\end{table}
The model coefficients are computed by the approach discussed in section \ref{sec:ord_noise_known}. We stack the measurements with $L = \hat{\eta} = 2$ and use the estimated error covariance matrix to compute the parameters.  The average of estimated model coefficients from 100 runs of MC simulations results in the following estimated ARX model.
\begin{align*}
y^{\star}[k] - \underset{\pm 0.03}{1.498}y^{\star}[k-1] + \underset{ \pm 0.03}{ 0.699}y^{\star}[k-2] = \\ -\underset{\pm 0.03}{0.001}u^{\star}[k] +
\underset{\pm 0.05}{0.995}u^{\star}[k-1] + \underset{\pm 0.05}{0.5}u^{\star}[k-2] 
\end{align*}
In the above model, we have also provided twice the standard deviation ($\sim$ 95 \% confidence interval) for each of the model coefficient estimates.  It can be observed that an unbiased estimate of the model parameters is obtained.  

The effect of different choices of $L$ and the number of measurements on the performance of the algorithm is further assessed.  
\subsubsection{Higher stacking order}
For the same example considered in the preceding section, we choose a higher value of stacking order $L = 15$ and assess the performance of our proposed algorithm. We perform the hypothesis test for equality of eigenvalues and the results are presented in Table \ref{tab:Ex1hypotest_L15}. 
\begin{table}[H]
\caption{Hypothesis test results of eigenvalues for $L = 15$}
    \label{tab:Ex1hypotest_L15}
    \begin{center}
    \begin{tabular}{c|c|c|c} \toprule
    $d_{\text{guess}}$ & Degrees of freedom ($\nu$) & Test statistic &  Test criterion ($\alpha$ = 0.05) \\ \midrule
    15 & 119 & 2581 & 139 \\
14 & 104 & 60.7 & 122.8 \\ 
\bottomrule
    \end{tabular}
    \end{center}
\end{table}

It can be inferred that the last $14$ eigenvalues are equal and hence the order can be estimated using $\hat{\eta} = L - \hat{d} + 1 = 15- 14 + 1 = 2$.  The model parameters estimated from 100 runs of MC simulations are reported in the difference equation given below:
\begin{align}
y^{\star}[k] - \underset{\pm 0.034}{1.498}y^{\star}[k-1] + \underset{ \pm 0.033}{ 0.699}y^{\star}[k-2] = \nonumber \\ \underset{\pm 0.034}{0.002}u^{\star}[k] + 
\underset{\pm 0.049}{0.991}u^{\star}[k-1] - \underset{\pm 0.054}{0.498}u^{\star}[k-2] 
\end{align}
It can be easily inferred that the estimated parameters are close to the true values used in simulation. In the next sub-section, we inspect the consistency of estimated parameters empirically. 
\subsubsection{Different sample sizes} In this subsection, we evaluate the algorithm's performance on the same example for different sample sizes $N = \{511,4095,8191\}$. The proposed algorithm is applied using $L = 5$ for all cases. The mean ($\mu$) and standard deviations ($\sigma$) of the last six converged eigenvalues obtained from 100 runs of MC simulations are presented in Table \ref{tab:ex1_sampl_eig}. It can be observed that the last $4$ eigenvalues are unity, and hence the order can be estimated correctly. 
\begin{table}[H]
	\caption{Eigenvalues for 100 MC simulations for different sample size} \label{tab:ex1_sampl_eig}
	\centering
	\begin{footnotesize}
		\begin{tabular}{c|c|c|c|c|c|c}
			\hline
			{}  &  \multicolumn{2}{|c|}{ N = 511} & \multicolumn{2}{|c|}{ N = 4095} & \multicolumn{2}{|c}{ N = 8191}\\
			\hline
			  & $\mu $ & $2\sigma$& $\mu $ & $2\sigma$  & $\mu $ & $2\sigma$ \\\hline
$\lambda_{7}$ &		9.782 &   2.537 &  11.056  &  0.83  & 11.126  &  0.818 \\
$\lambda_{8}$ &     5.46 &   0.884 &   5.768  &  0.305  &  5.759  &  0.28 \\
$\lambda_{9}$ &     1.07 &   0.07 &   1.03  &  0.032  &  1.026  &  0.033 \\
$\lambda_{10}$ &    1.025 &   0.071 &   1.012  &  0.035  &  1.007  &  0.035 \\
$\lambda_{11}$ &    0.961 &   0.086 &   0.984  &  0.04  &  0.983  &  0.035 \\
$\lambda_{12}$ &    0.908 &   0.101 &   0.958  &  0.05  &  0.962  &  0.052 \\
			\hline
		\end{tabular}
	\end{footnotesize} 
\end{table}
\vskip -0.35cm
The mean and the standard deviations of the estimated parameters are reported  in Table \ref{tab:mcsim_N}. 
\begin{table}[H]
	\caption{MC simulation results for various sample sizes \label{tab:ex1_sampl}}
	\label{tab:mcsim_N}
	\centering
	\begin{footnotesize}
		\begin{tabular}{cc|c|c|c|c|c|c}
			\hline
			{} & {} &  \multicolumn{2}{|c|}{ N = 511} & \multicolumn{2}{|c|}{ N = 4095} & \multicolumn{2}{|c}{ N = 8191}\\
			\hline
			  &Value& $\mu $ & $2\sigma$& $\mu $ & $2\sigma$  & $\mu $ & $2\sigma$ \\\hline
			$\sigma^2_{e_y}$ &0.2&0.217  &  0.054& 0.21   & 0.012 &0.207 & 0.007\\
			$\sigma^2_{e_u}$ &0.1& 0.084 &   0.048 & 0.0913&    0.014 & 0.0939&    0.011\\
			$a_1$ &-1.5& -1.499 &  0.037 & -1.496 & 0.024 & -1.497 & 0.021 \\ 
			$a_2$ &0.7& 0.7   &  0.035  & 0.697 & 0.022  & 0.698 & 0.02 \\
			$b_0$ &0  & 0.003 &  0.047   & 0.001 &  0.017 & 0 & 0.013\\ 
			$b_1$ &1& 0.994  & 0.068   & 0.999 & 0.03  & 0.997 & 0.023 \\ 
			$b_2$ &0.5  & 0.492 &  0.073  & 0.504  & 0.04 & 0.501 & 0.035 \\ 
			\hline
		\end{tabular}
	\end{footnotesize} 
\end{table}
It can be observed that unbiased estimates of model parameters are obtained in all cases. It can also be noted that the standard deviation of errors in the estimated parameters decreases with increasing sample size, indicating that perhaps the proposed method also delivers consistent estimates. 
\subsection{Example 2: SISO with system order 3 and input dynamics}
\label{ex:ex2}
\begin{figure}[thpb]
      \centering
      \includegraphics[scale=0.18]{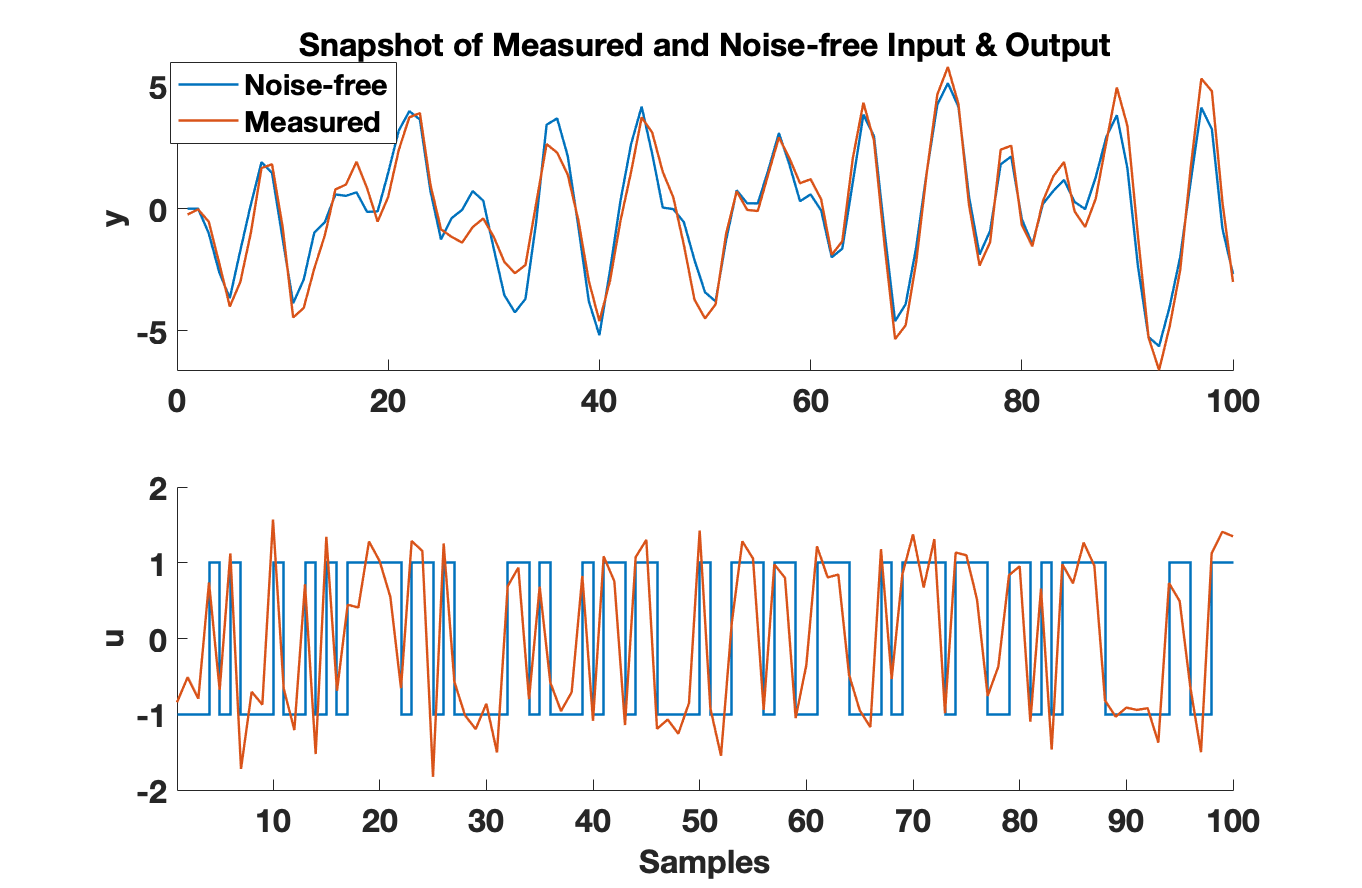}
      \caption{A snapshot for the input and output data in Example \ref{ex:ex2}}
      \label{fig:iodata_ex2}
\end{figure}

In this subsection, we evaluate the proposed method on a SISO system with system order ($\eta$) of $3$. 
\begin{align}
    y^{\star}[k] -1.1y^{\star}[k-1] + 0.7y^{\star}[k-2] = \nonumber \\  u^{\star}[k-2] + 0.5 u^{\star}[k-3]
    \label{eq:ex2_diffeqn}
\end{align}
The system is simulated to generate  $4095$ noisy measurements of input and output data. The input and output noise variances were chosen to be $0.1$ (= $\sigma^2_{e_u}$) and $0.15$ (= $\sigma^2_{e_y}$), respectively, to maintain SNR of 10.  A snapshot of a few samples of measured input-output samples is shown in Figure \ref{fig:iodata_ex2}. It should be noted that the output noise follows the ARX model with parameters described in \eqref{eq:ex2_diffeqn}.

The lagged data matrix $\mathbf{Z}_L$ is constructed for $L = 6$, and the proposed algorithm is applied starting with the maximum possible value of $d_{\text{guess}} = 6$.  We apply the hypothesis test as described in \cite{mann2020optimal} to test whether the smallest $d_{\text{guess}}$ eigenvalues are all equal. The hypothesis test results are presented in Table \ref{tab:Ex2hypotest} for different values of $d_{\text{guess}}$.

\begin{table}[H]
\caption{Hypothesis test results for eigenvalues in Example \ref{ex:ex2}}
    \label{tab:Ex2hypotest}
    \begin{center}
    \begin{tabular}{c|c|c|c} \toprule
    $d_{\text{guess}}$ & Degrees of freedom ($\nu$) & Test statistic &  Test criterion ($\alpha$ = 0.05) \\ \midrule
    6 & 20 & 1.6 $\times 10^4$ & 28.4 \\
5 & 14 & 539.7 & 21.1  \\ 
4 & 9 & 1.1 & 14.7 \\ 
\bottomrule
    \end{tabular}
    \end{center}
\end{table}
It can be observed that the null hypothesis is not rejected for $d_{\text{guess}} = 4$.  Therefore, the system order is correctly estimated as $\hat{\eta} = L - \hat{d} + 1 = 6-4+1= 3$.  We use the estimated system order and error variances to obtain the model parameters as described in section  \ref{sec:ord_noise_known}.  The mean ($\mu$) and standard deviation ($\sigma$) of the estimated model parameters from 100 runs of MC simulations are shown in the last two columns of Table \ref{tab:ex2_comp}. It can be verified that unbiased estimates of all parameters are obtained.
We also compare the results obtained using our proposed algorithm with prediction error method used for classical ARX model identification, which assumes that the input is noise free.  In this case, if the model order is specified, the ARX model parameters can be obtained using ordinary least squares. The parameter estimates obtained using this method are presented in the first two columns of Table \ref{tab:ex2_comp}. 
\begin{table}[H]
	\caption{MC simulation results for 100 runs}
	\label{tab:ex2_comp}
	\centering
	\begin{footnotesize}
		\begin{tabular}{cc|c|c|c|c}
			\hline
			\multicolumn{2}{c|}{Model $\rightarrow$} &   \multicolumn{2}{|c|}{ARX} & \multicolumn{2}{|c}{ARX-EIV }\\
			\hline
			\multicolumn{2}{c|}{Algorithm $\rightarrow$} &   \multicolumn{2}{|c|}{Least Squares} & \multicolumn{2}{|c}{Proposed }\\
			\hline
			  &Value$\downarrow$&$\mu $ & $2\sigma$  & $\mu $ & $2\sigma$ \\\hline
	$\sigma^2_{e_y}$ & 0.15 & 0.26 & 0.012 & 0.14 & 0.026\\
	$\sigma^2_{e_u}$ & 0.1 & - & - & 0.106  &  0.02 \\
			$a_1$ &-1.1& -1.12  &  0.01 & -1.072 & 0.065\\ 
			$a_2$ &0.7 & 0.712  &  0.01  & 0.67 & 0.064 \\
			$a_3$ &0 & -  & -  & 0 & 0.055 \\
			$b_0$ &0 & -  & -  & 0 & 0.02 \\
			$b_1$ &0  &  - & - &  0.01 & 0.025\\ 
			$b_2$ &1&  0.91 & 0.015 &  0.99 & 0.045\\ 
			$b_3$ &0.5&  0.44 & 0.02 &  0.52 & 0.07\\ 
			\hline
		\end{tabular}
	\end{footnotesize} 
\end{table}
It can be deduced that ignoring input noise and directly applying the least-squares method leads to biased estimates (as expected). It should also be noted that correct values of input-output orders and delay were supplied to the least-squares algorithm, unlike the proposed algorithm which is not provided any such information. 

In order to assess the performance of the proposed method when the noise in input is low, the same system is simulated with input noise variance $\sigma^2_{e_u} = 0.01$.  The results of applying the least squares method and proposed approach are presented in Table \ref{tab:ex2_comp_inpsnrhigh}. 
\begin{table}[H]
	\caption{MC simulation results for at high SNR for input}
	\label{tab:ex2_comp_inpsnrhigh}
	\centering
	\begin{footnotesize}
		\begin{tabular}{cc|c|c|c|c}
			\hline
			\multicolumn{2}{c|}{Model $\rightarrow$} &   \multicolumn{2}{|c|}{ARX} & \multicolumn{2}{|c}{ARX-EIV }\\
			\hline
			\multicolumn{2}{c|}{Algorithm $\rightarrow$} &   \multicolumn{2}{|c|}{Least Squares} & \multicolumn{2}{|c}{Proposed }\\
			\hline
			  &Value$\downarrow$&$\mu $ & $2\sigma$  & $\mu $ & $2\sigma$ \\\hline
	$\sigma^2_{e_y}$ & 0.15 & 0.16  &  0.007   & 0.136  &  0.021\\
	$\sigma^2_{e_u}$ & 0.01 & - & - & 0.018   & 0.016 \\
			$a_1$ &-1.1& -1.1  &  0.01& -1.071   & 0.046\\ 
			$a_2$ &0.7 & 0.701 & 0.01  & 0.673   & 0.051\\
			$a_3$ &0 & -  & -  & 0.001 &   0.036 \\
			$b_0$ &0 & -  & -  & 0 & 0.02 \\
			$b_1$ &0  &  - & - &  0.007  &  0.018\\ 
			$b_2$ &1&  0.991 & 0.011 &  0.991  &  0.029\\ 
			$b_3$ &0.5&  0.493  & 0.017 &  0.524  &  0.049\\ 
			\hline
		\end{tabular}
	\end{footnotesize} 
\end{table}
In this case, the least squares approach gives unbiased estimates of the parameters, indicating that the input noise is negligible.  Although, the proposed method also provides unbiased estimates of all model parameters, the estimates are less accurate as compared to the estimates obtained using least squares.  This can be attributed to the fact that in the proposed method more parameters are estimated, since no knowledge of the system is assumed.  As an extension, it is also possible to examine the estimated input and output error variances obtained using the proposed approach and make a decision whether to ignore the input noise and accept the least squares estimates, especially if the output error variance is one or two orders of magnitude higher than the input error variance. Even in this case, the proposed approach provides valuable information regarding the system order and the significant model coefficients that need to be estimated using the least squares approach. 


\section{Conclusion}
\label{sec:conc}
In this paper, we propose a novel identification algorithm capable of estimating the order, delay, noise variances, and coefficients of an EIV-ARX model for SISO systems.  Unlike existing methods, no prior knowledge of system parameters is required in our proposed procedure. Simulation studies indicate that the estimated parameters are unbiased and consistent, although this has not been theoretically established. Extension of the proposed work for identifying other model structures such as ARMAX models may be considered in the future. 

\bibliographystyle{IEEEtran}
\bibliography{main}
\end{document}